\documentclass[times,twocolumn,final,authoryear]{elsarticle}

\usepackage{jasr}
\usepackage{framed,multirow}
\usepackage{amsmath,amssymb}
\usepackage{latexsym}
\usepackage{xcolor}
\usepackage[citebordercolor=white]{hyperref}
\usepackage{graphicx}
\usepackage{enumitem}
\usepackage[switch]{lineno}

\journal{Advances in Space Research}

\begin{document}

\verso{Borka \textit{et al}}

\begin{frontmatter}

\title{Velocity distribution of elliptical galaxies in the framework of Non-local Gravity model}

\author[1]{Du\v{s}ko \snm{Borka}\corref{cor1}}
\cortext[cor1]{Corresponding author.}
\ead{dusborka@vinca.rs}
\author[1]{Vesna \snm{Borka Jovanovi\'{c}}}
\ead{vborka@vinca.rs}
\author[2,3]{Salvatore \snm{Capozziello}}
\ead{capozzie@na.infn.it}
\author[4]{Predrag \snm{Jovanovi\'{c}}}
\ead{pjovanovic@aob.rs}

\address[1]{Department of Theoretical Physics and Condensed Matter Physics (020), Vin\v{c}a Institute of Nuclear Sciences - National Institute of the Republic of Serbia, University of Belgrade, P.O. Box 522, 11001 Belgrade, Serbia}
\address[2]{Dipartimento di Fisica ''E. Pancini'', Universit\`{a} di Napoli ''Federico II'', Compl. Univ. di Monte S. Angelo, Edificio 6, Via Cinthia, I-80126, Napoli, Italy}
\address[3]{Scuola Superiore Meridionale, Largo S. Marcellino 10, I-80138, Napoli, Italy}
\address[4]{Astronomical Observatory, Volgina 7, P.O. Box 74, 11060 Belgrade, Serbia}

\received{}
\finalform{}
\accepted{}
\availableonline{}
\communicated{}

\begin{abstract}
We investigate the velocity distribution of elliptical galaxies in the framework of Non-local Gravity. According to this approach, it is possible to  recover the fundamental plane  of elliptical galaxies  without the dark matter hypothesis. Specifically, we compare theoretical predictions for circular velocity in Non-local Gravity context with the corresponding values coming from a large sample of observed elliptical galaxies. We adopt the surface brightness, effective radius and velocity dispersion as structural parameters for the fundamental plane. As final result, it is possible to show  that non-local gravity effects can reproduce the stellar dynamics in elliptical galaxies and fit consistently observational data.
\end{abstract}

\begin{keyword}
\KWD Modified  gravity \sep Elliptical galaxies \sep Fundamental plane
\end{keyword}

\end{frontmatter}

%% For linenumbers
%\linenumbers

%% main text

\section{Introduction}

Velocity distribution of elliptical galaxies is an important physical quantity to  recover the fundamental plane (FP) of elliptical galaxies. The FP is an empirical relation among three global parameters of elliptical galaxies expressed as a relationship between the central projected velocity dispersion $\sigma_0$, the effective radius $r_e$, and the mean effective surface brightness (within $r_e$) $I_e$ \citep{dres87,ciot97}. Elliptical galaxies are confined in a narrow logarithmic plane of their configuration space dubbed FP \citep{dres87,ciot97}. In this perspective, it is worth noticing that several  features of  galaxies are correlated. For example,  a galaxy with a higher luminosity has a larger effective radius. The relevant result is that in any case, elliptical galaxies and any spheroidal self-gravitating, collisionless system (e.g. the bulge of spiral galaxies) lay on the related FP.  The features of FP are defined and discussed in detail in several papers, see e.g \citet{gude73,gude91,bend92,bend93,busa97,binn98,saul13,tara15,terl81} and references therein. The FP empirical relation is given by the following equation \citep{busa97}:

\begin{equation}
\log(r_e) = a \, \log(\sigma_0) + b \, \log(I_e) + c,
\label{equ01}
\end{equation}

\noindent with $a$ and $b$ being the FP coefficients which are fixed by observations. This relation gives us the possibility to obtain observational constraints on the structure, formation, and evolution of early-type galaxies and, in general, on spheroidal,  self-gravitating  systems. Reversing the argument, the FP can be adopted to fix parameters of a given theory of gravity, once they are constrained by observations.

In this paper, we shall constrain parameters of Non-local Gravity with the aim to show that we do not need any dark matter to fix the FP structure. In order to describe the velocity of stellar populations, one can define rotational velocity of a group of stars $v_c$ and dispersion $\sigma$ which represents the characteristic random velocity of stars. Then, the obtained ratio $v_c/ \sigma$ is a relation which characterizes the kinematics of galaxies. In case of spiral galaxies, the  ratio is $v_c/ \sigma \gg 1$ and it represents  kinematically "cold systems", while elliptical galaxies are characterized by $0 < v_c/ \sigma < 1$ and it represents kinematically "hot systems".

Our aim is to investigate velocity distribution $\sigma$ of elliptical galaxies as an important physical quantity for recovering the empirical relation named fundamental plane (FP) of elliptical galaxies. In the case of the spirals, FP does not hold, so it is more appropriate to test rotation galactic curves \citep{capo07}.

Here, we adopt FP to constrain Non-local Gravity, a particular class of Extended Theories of Gravity using astronomical observations for velocity distributions of elliptical galaxies.  Extended Gravity is presented in several review papers like \citet{capo11,noji11,noji17,capo13,capo19}. Some experimental limits related to Extended Theories of Gravity are reported in \citet{avil12,duns16,tino20,capo15} and references therein.

The layout of the  paper is the following. In Section 2, we briefly sketch Non-local Gravity. In Section 3, we present the singular isothermal sphere model and velocity distribution in relation to Non-local Gravity. Adopted data are presented in   Section 4. In Section 5, we find constraints on Non-local Gravity parameters using astronomical data and we study the velocity distribution of elliptical galaxies. Section 6 is devoted to summarize the results and to draw  conclusions.

\section{Non-local Gravity as a link between Quantum Mechanics and General Relativity}

The main feature of Quantum Mechanics is related to the Heisenberg Principle which points out indetermination, and then non-locality, in physics. On the other hand, General Relativity  is a genuinely local description of gravitational field. Adding non-local terms into the Einstein-Hilbert action can constitute a link to join the two main theories of modern physics. The non-locality can be straightforwardly introduced by the inverse of the d'Alembert operator. In Refs. \citet{koiv08a,koiv08b,barv15,capo21a,capo20b}, a discussion on dynamics of Non-local Gravity is reported. In particular, its Newtonian limit is taken into account.
It was recently showed that Non-local Gravity can suitably represent the behavior of gravitational interaction without the need for dark matter and dark energy at different astrophysical and cosmological scales. 

A possible action describing  Non-local Gravity was suggested in \citet{dese07}. The authors proposed a non-local modification of  Einstein-Hilbert action with the following form:

\begin{equation}
\mathcal{S} =\frac{1}{2 \kappa ^2} \int d^4 x \sqrt{-g} \left[ R 
\left( 1 + f(\square ^{-1} R) \right) \right] \,
\label{equ02}
\end{equation}
where $R$ is the Ricci scalar and $f(\square^{-1}R)$ is an arbitrary function of the non-local term $\square ^{-1}R$. This non-local term is called \textit{distortion function} and it is  given by a retarded Green's function of the form:

\begin{equation}
\mathcal{G}[f](x)=(\square ^{-1}f)(x) = \int d^4x' \sqrt{-g(x')}f(x')G(x,x')\,.
\label{equ03}
\end{equation}
If we set $f(\square^{-1}R)=0$,  the Einstein theory is immediately recovered.  It is possible to show that the weak-field limit of this Non-local Gravity model has the free parameters, $\phi_c, \, r_{\phi}$ and $r_{\xi}$ (see \citet{dial19} for more details). We can reasonably take specific values for $\phi_c$ around  1 (from 0.9 to 1.1), and constrain the parameter space of $r_{\phi}$ and $r_{\xi}$ parameters. The related weak field potential reads:

\begin{eqnarray}
\Phi(r) = &-& \frac{G M}{r}\phi_c\ + \frac{G^2 M^2}{2c^2r^2} \left[\frac{14 }{9}\phi _c^2 + \frac{18 r_{\xi }-11 r_{\phi }}{6 r_{\xi} r_{\phi}} r \right] +  \nonumber\\
&+& \frac{G ^3 M^3}{2c^4r^3} \left[\frac{7 r_{\phi }-50 r_{\xi }}{12 r_{\xi } r_{\phi }} \phi _c r-\frac{16 \phi_c^3}{27} + \frac{2 r_{\xi }^2-r_{\phi }^2}{r_{\xi }^2 r_{\phi}^2}r^2\right]\,. \nonumber\\
\label{equ04}
\end{eqnarray}
where $G$ is the Newtonian constant, $M$ is the mass of the system generating the gravitational field, and $\phi_c$ is a dimensionless constant. Two new length scales,  denoted by $r_{\phi}$ and $r_{\xi}$, arise in relation  to the two scalar degrees of freedom, $\phi$ and $\xi$  respectively. These scalar fields represent the non-locality \citep{capo20b,capo21a,capo21b,acun22,capo22}.

In paper \citet{noji20} the authors demonstrate that the non-local $f(R)$ theories of gravity can be compromised by the existence of ghost degrees of freedom, but in addition they proposed a direct remedy for the non-local $f(R)$ gravity, by providing a ghost-free modification of the non-local $f(R)$ gravity theory. In case of non-local gravity model that we used [\citet{dial19}, equation (12)], we do not get ghosts in the parameter interval we studied. More details of ghost free study can be find in \citet{noji20}, \citet{noji21} and in references therein.

\section{The singular isothermal sphere model and velocity distributions in Non-local Gravity}

For modeling the stellar kinematics, i.e. in order to describe the mass distribution in elliptical galaxies, we assume that the mass distribution within them is described by the singular isothermal sphere (SIS) model: $M(r) = 2\sigma_{SIS}^2\,G^{-1}\,r$, like in our previous paper \citep{capo20a}. The density profile has the form: $\rho_{SIS}(r) = \dfrac{\sigma_{SIS}^2}{2\pi Gr^2}$, and the corresponding mass within a radius $r$ is: 

\begin{equation}
M_{SIS}(r) = \dfrac{2\sigma_{SIS}^2}{G}\cdot r.
\label{equ05}
\end{equation}
We can see that mass grows linearly with $r$. Taking into account: 

\begin{equation}
v_N^{2}(r) = \dfrac{GM(r)}{r}
\label{equ06}
\end{equation}
and 
\begin{equation}
v_N(r_e) = \sigma_0,
\label{equ07}
\end{equation}

\noindent where $v_N(r_e)$ is the Newtonian circular velocity at the effective radius, and $\sigma_0$ is the observed velocity dispersion \citep{burs97}, we can obtain:

\begin{equation}
v_N^2(r) = 2\sigma_{SIS}^2.
\label{equ08}
\end{equation}
Therefore, for $r=r_e$ it stands:

\begin{equation}
\sqrt{2}\sigma_{SIS} = \sigma_0,
\label{equ09}
\end{equation}

\noindent and furthermore: 
\begin{equation}
v_N(r_e)^2 = \sigma_0^2.
\label{equ10}
\end{equation}
What we want to show here is that Non-local Gravity parameters can be constrained by the velocity distribution of elliptical galaxies.

To start our analysis, let us write the Newtonian potential in the form $\Phi_N(r) = -\dfrac{GM(r)}{r}$ and circular velocity as $v_N^{2}(r) = r \cdot \Phi_N^\prime(r)$. If we suppose that mass is spherically distributed in elliptical galaxies, we obtain the circular velocity $v_c^{2}(r) = r \cdot \Phi^\prime(r)$,  according to Eq.\eqref{equ04} for the Non-local Gravity potential. We are using following substitutions:

\begin{eqnarray}
v_N^{2}(r) = \dfrac{GM(r)}{r}, \nonumber\\
\Phi_N(r) = -\dfrac{GM(r)}{r}, \nonumber\\
\Phi_N(r)^\prime = (-\dfrac{GM(r)}{r})^\prime, \nonumber\\
v_N^{2}(r) = -\Phi_N(r)^\prime.
\label{equ11}
\end{eqnarray}
In order to derive expression for velocity distribution, we start from the non-local gravitational potential and derive the connection between $v_c^{2}(r)$ and parameters of this potential, we obtain:

\begin{eqnarray}
v_c^2(r) &=& \phi_c \, v_N^2(r) - \dfrac{v_N^4(r)}{c^2} \left[ \dfrac{r}{2} \cdot \dfrac{18 r_\xi - 11 r_\phi}{6 r_\xi \, r_\phi} + \dfrac{14}{9}\phi_c^2 \right] + \nonumber \\
&+& \dfrac{v_N^6(r)}{2 c^4} \left[-\dfrac{\phi_c r}{6} \cdot \dfrac{7r_\phi - 50 r_\xi}{r_\xi \, r_\phi} + \dfrac{16}{9} \phi_c^3 - r^2 \dfrac{2 r_\xi^2 - r_\phi^2}{r_\xi^2 \, r_\phi^2} \right]. \nonumber  \\
\label{equ12}
\end{eqnarray}

If we take into account that the Newtonian circular velocity at the effective radius for elliptical galaxies, i.e. for $r = r_e$, is $v_N(r_e) = \sigma_0$, where $\sigma_0$ is the observed velocity dispersion \citep{burs97}, the velocity dispersion becomes:

\begin{equation}
\sigma^{theor}(r_e) = v_c(r_e) = \sigma_0.
\label{equ13}
\end{equation}

This is not the case for non-ellipticals. After combining Eq.\eqref{equ12} and Eq.\eqref{equ13} we obtained:

\begin{eqnarray}
	1 &=& \phi_c \ - \dfrac{\sigma_0^2}{c^2} \left[ \dfrac{r_e}{2} \cdot \dfrac{18 r_\xi - 11 r_\phi}{6 r_\xi \, r_\phi} + \dfrac{14}{9}\phi_c^2 \right] + \nonumber \\
	&+& \dfrac{\sigma_0^4}{2 c^4} \left[-\dfrac{\phi_c r_e}{6} \cdot \dfrac{7r_\phi - 50 r_\xi}{r_\xi \, r_\phi} + \dfrac{16}{9} \phi_c^3 - r^2_e \dfrac{2 r_\xi^2 - r_\phi^2}{r_\xi^2 \, r_\phi^2} \right]. \nonumber  \\
	\label{equ14}
\end{eqnarray}

From this expression, we can start our analysis.

\begin{figure*}[ht!]
\centering
\includegraphics[width=0.48\textwidth]{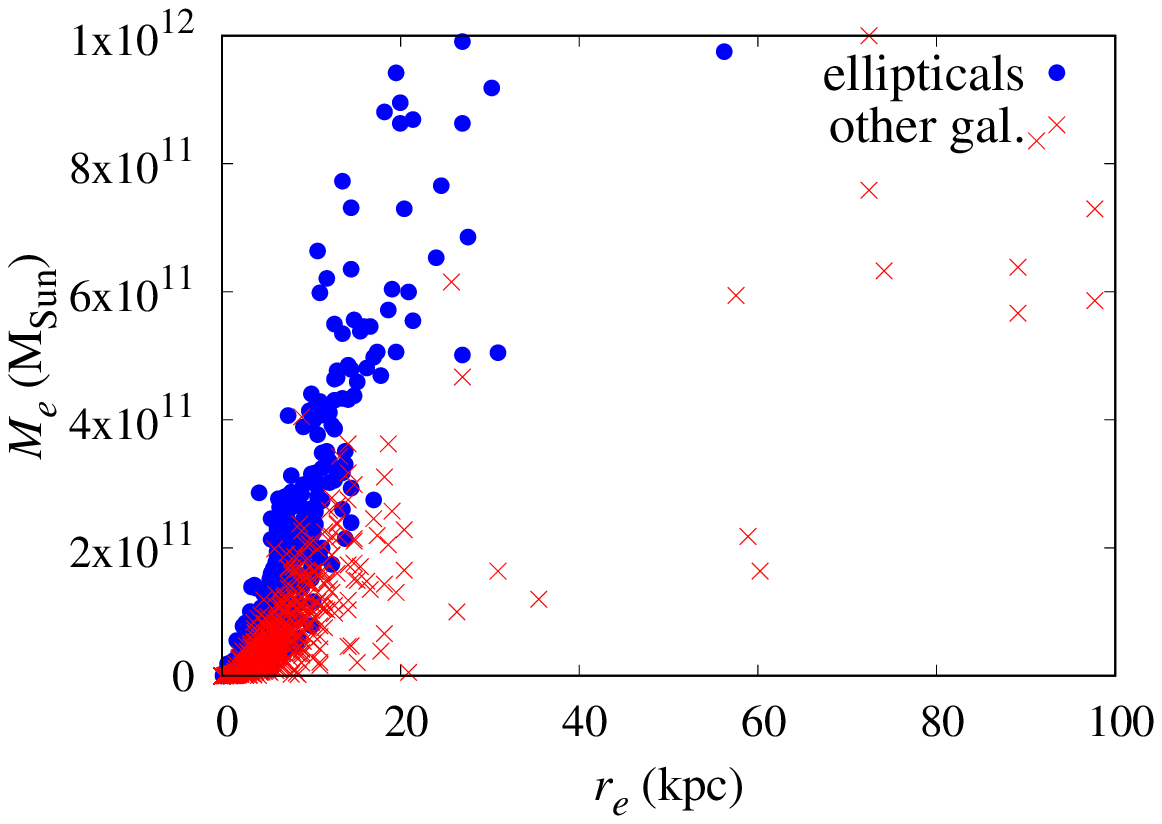}
\hfill
\includegraphics[width=0.48\textwidth]{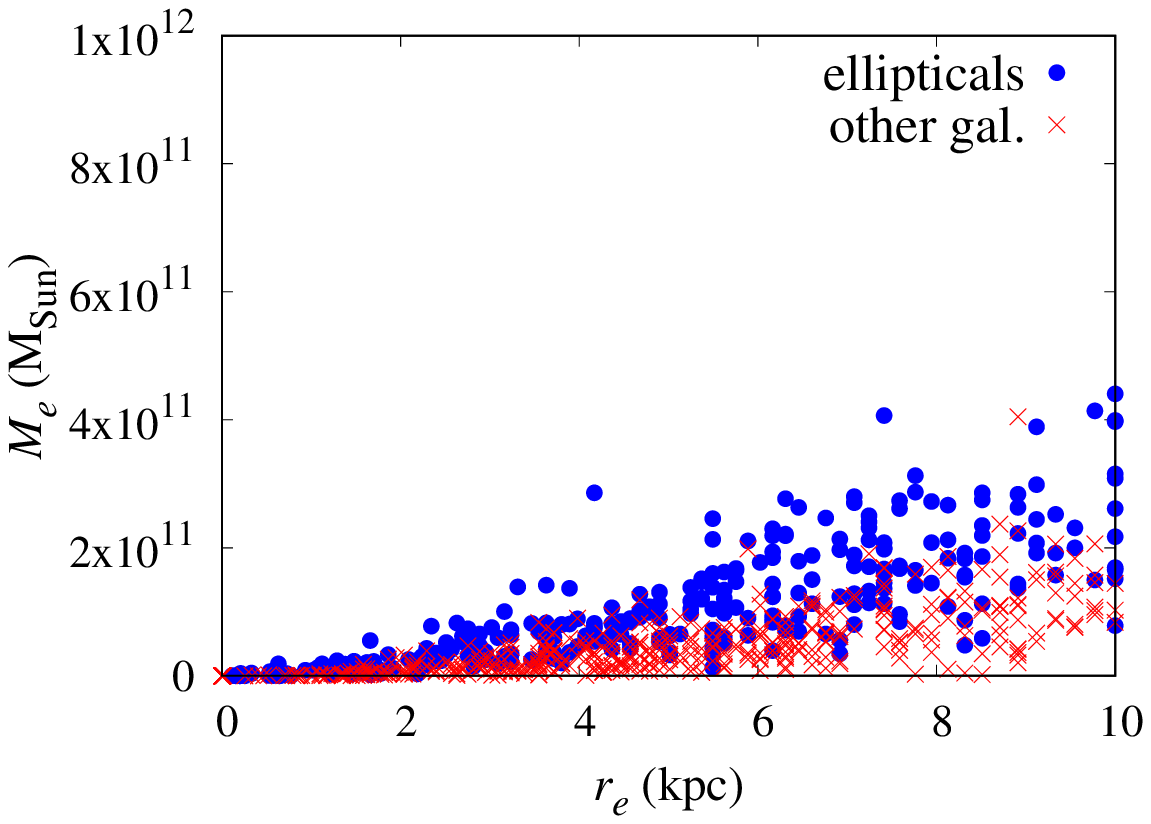}
\caption{(\textbf{left}): Galaxy masses $M_e$ as a function of effective radius $r_e$, for elliptical and other types of galaxies (for the sample of galaxies listed in Table 1 from \citet{burs97}). (\textbf{right}): Right panel shows a zoomed part of the figure, for $r_e$ less than 10 kpc.}
\label{fig01}
\end{figure*}

\begin{figure*}[ht!]
\centering
\includegraphics[width=0.48\textwidth]{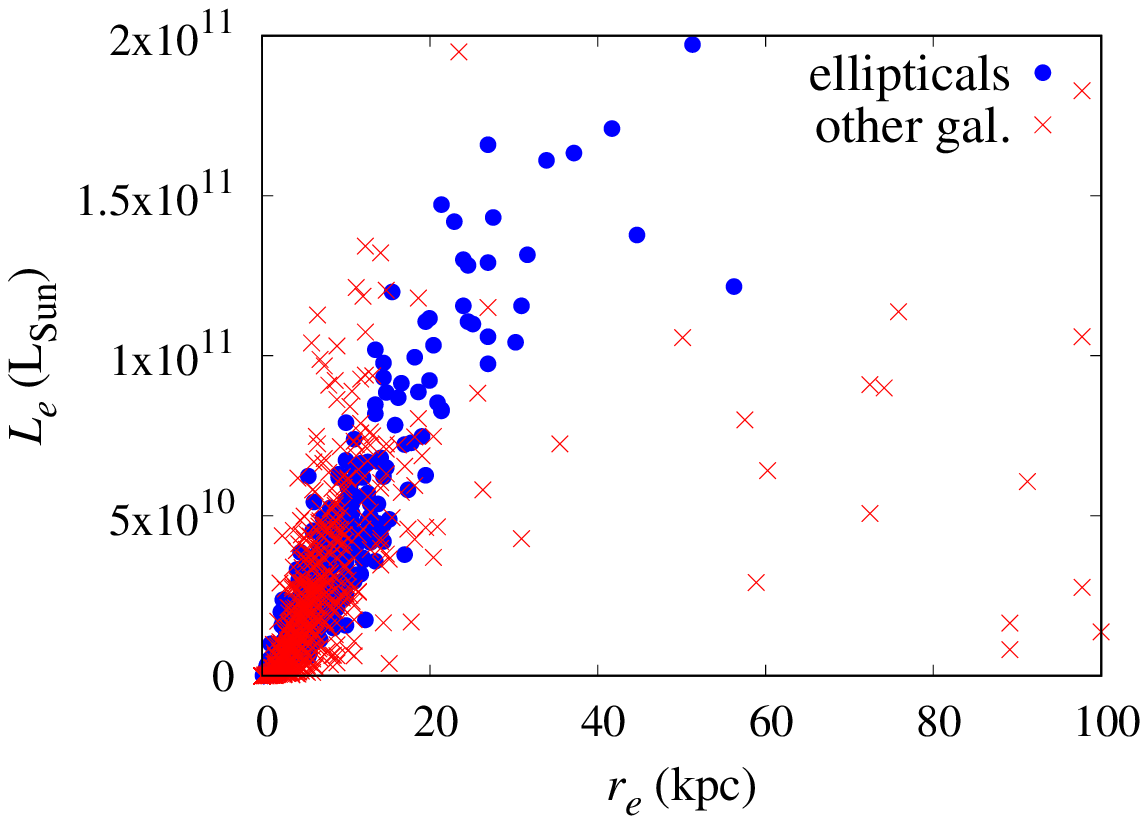}
\hfill
\includegraphics[width=0.48\textwidth]{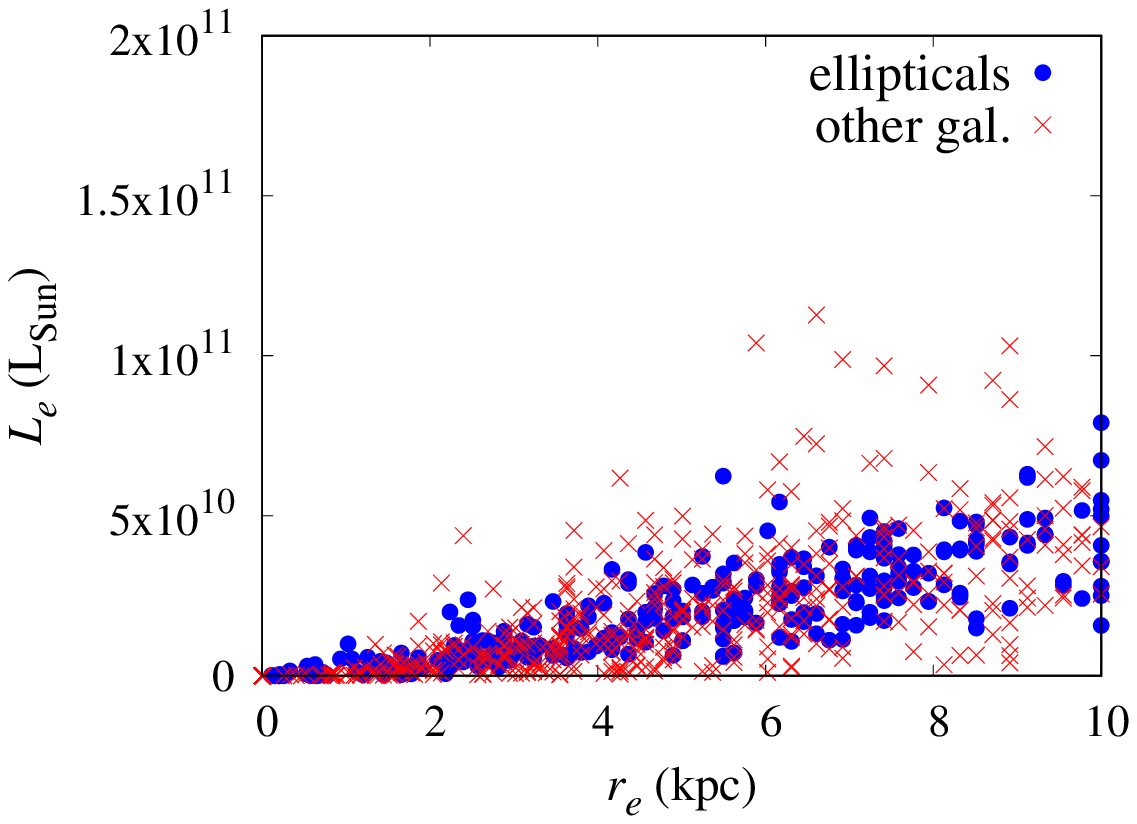}
\caption{(\textbf{left}): Galaxy luminosities $L_e$ as a function of effective radius $r_e$, for elliptical and other types of galaxies. (\textbf{right}): Right panel shows a zoomed part of the figure, for $r_e$ less than 10 kpc. Data are from \citet{burs97}.}
\label{fig02}
\end{figure*}

\begin{figure*}[ht!]
\centering
\includegraphics[width=0.82\textwidth]{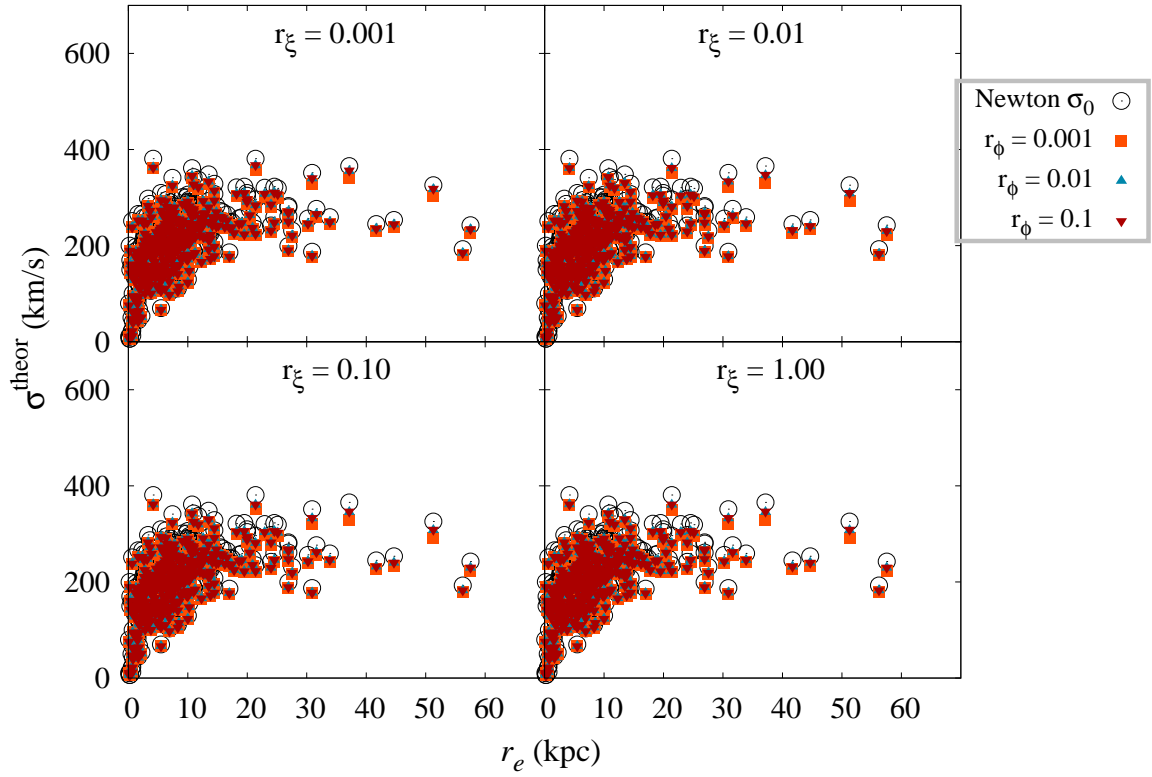}
\caption{Velocity dispersion $\sigma^{theor}$ as a function of the effective radius $r_e$ for elliptical galaxies, for four different values of the $r_{\xi}$: 0.001, 0.01, 0.10 and 1.00 kpc. The Newtonian velocity dispersion $\sigma_0$ at the effective radius  is taken from \citet{burs97}. Theoretical values of velocity dispersion $\sigma^{theor}$ are calculated for the three values of non-local Gravity parameter $r_{\phi}$: 0.001, 0.01 and 0.1 kpc. Value of $\phi_c$ is 0.9.}
\label{fig03}
\end{figure*}

\begin{figure*}[ht!]
\centering
\includegraphics[width=0.82\textwidth]{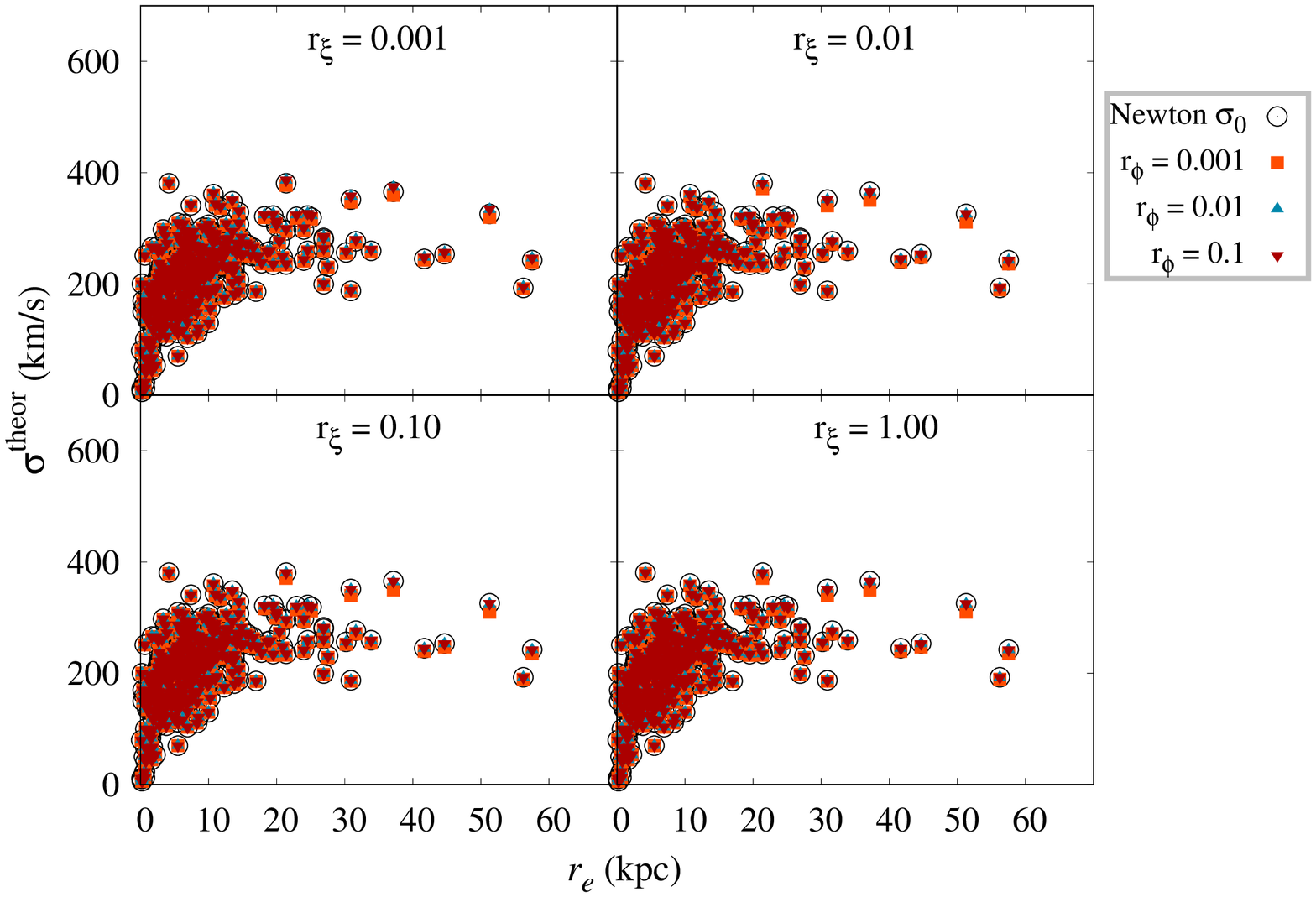}
\caption{The same as in Fig. \ref{fig03}, but for the value of $\phi_c$ equals 1.}
\label{fig04}
\end{figure*}

\begin{figure*}[ht!]
\centering
\includegraphics[width=0.82\textwidth]{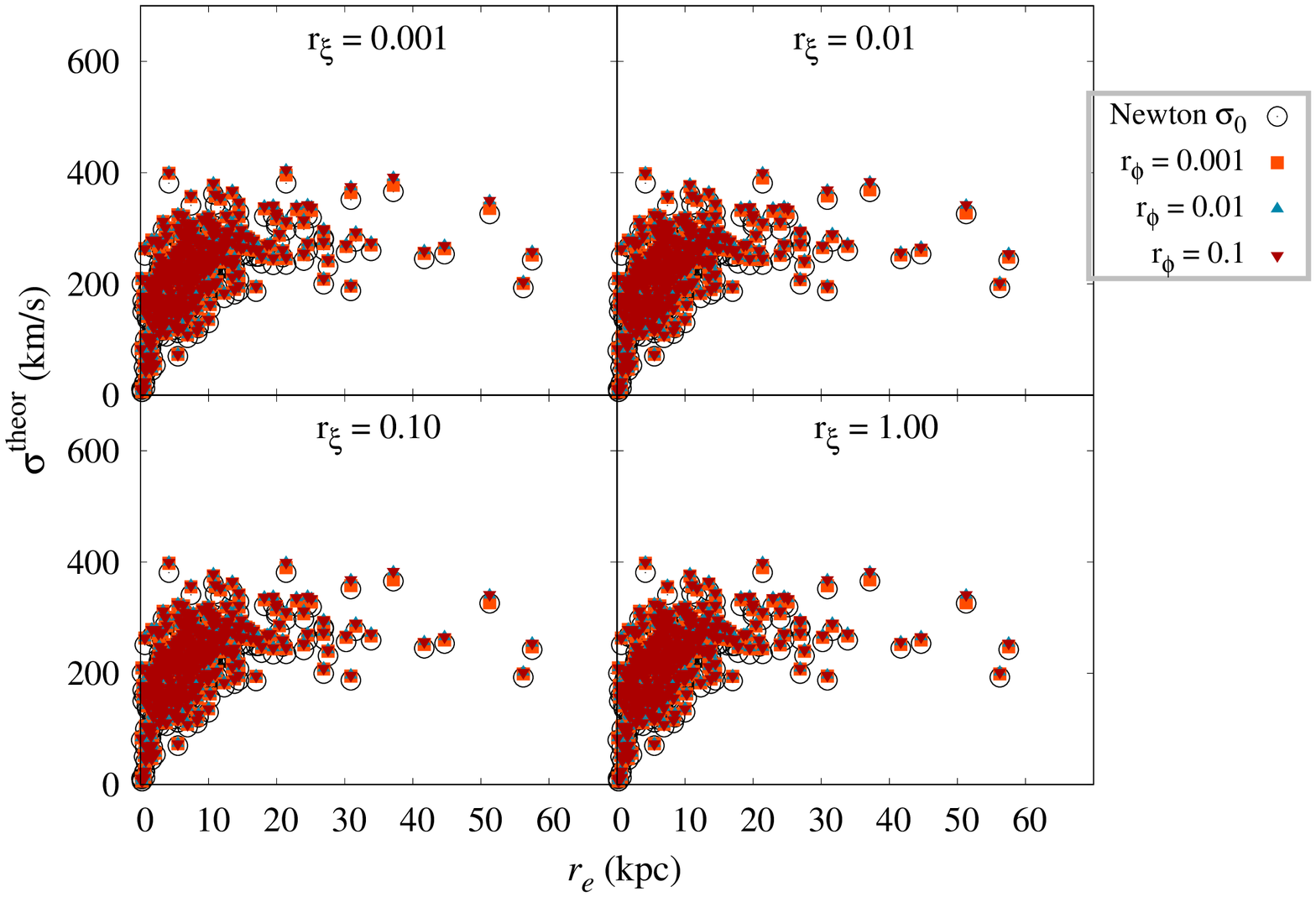}
\caption{The same as in Fig. \ref{fig03}, but for the value of $\phi_c$ equals 1.1.}
\label{fig05}
\end{figure*}

\begin{figure*}[ht!]
\centering
\includegraphics[width=0.45\textwidth]{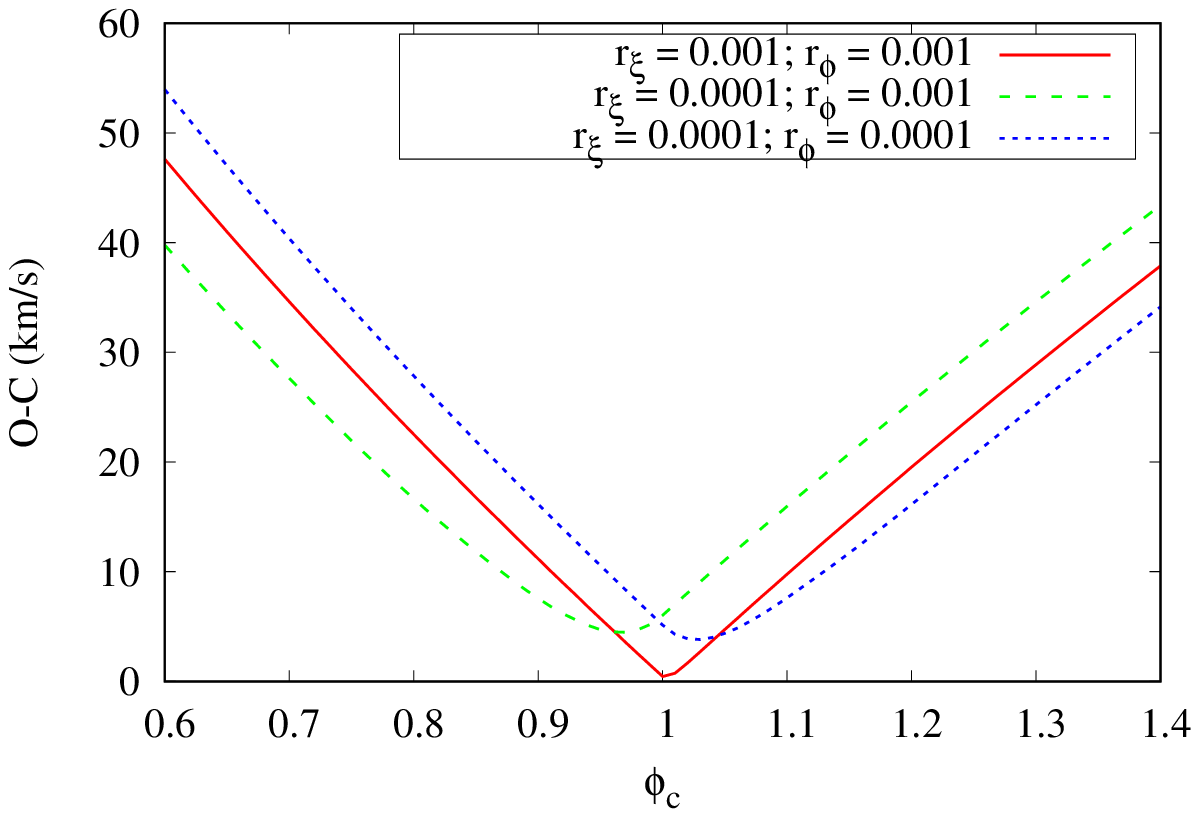}
\includegraphics[width=0.45\textwidth]{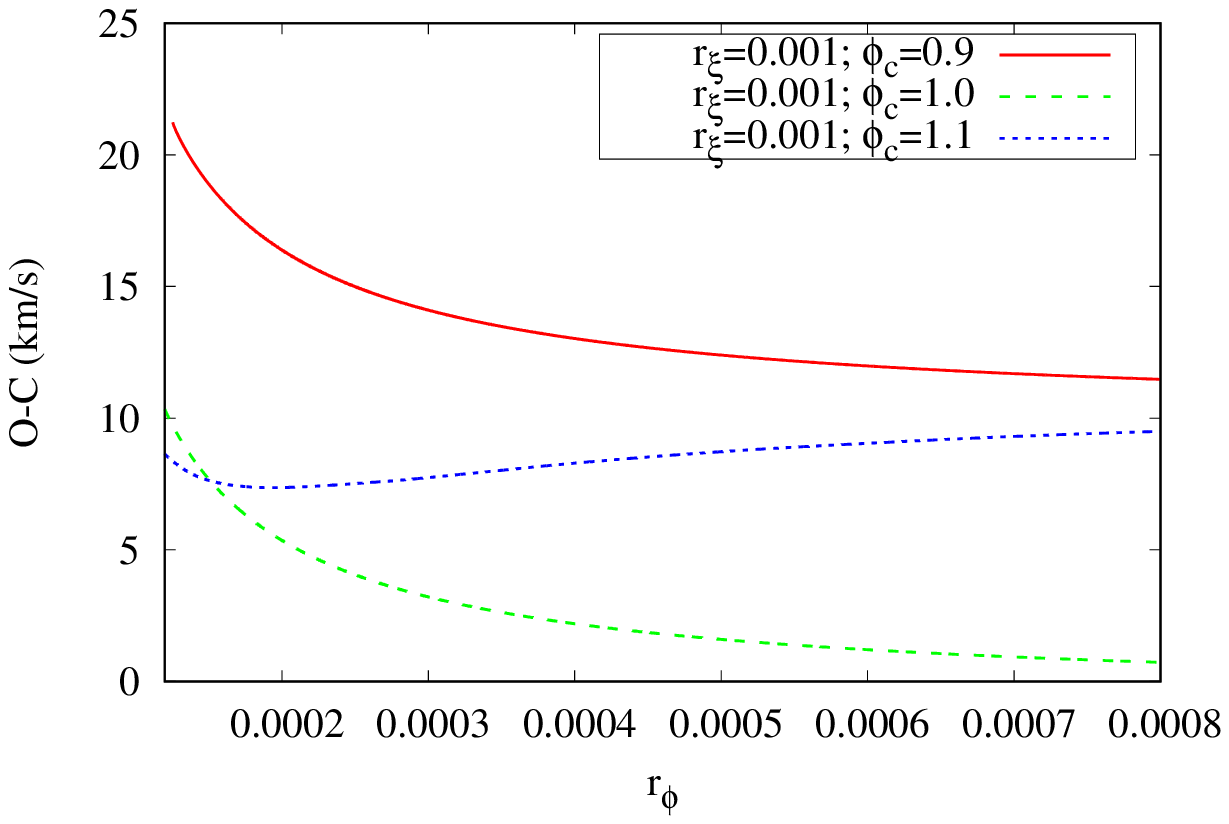}
\includegraphics[width=0.45\textwidth]{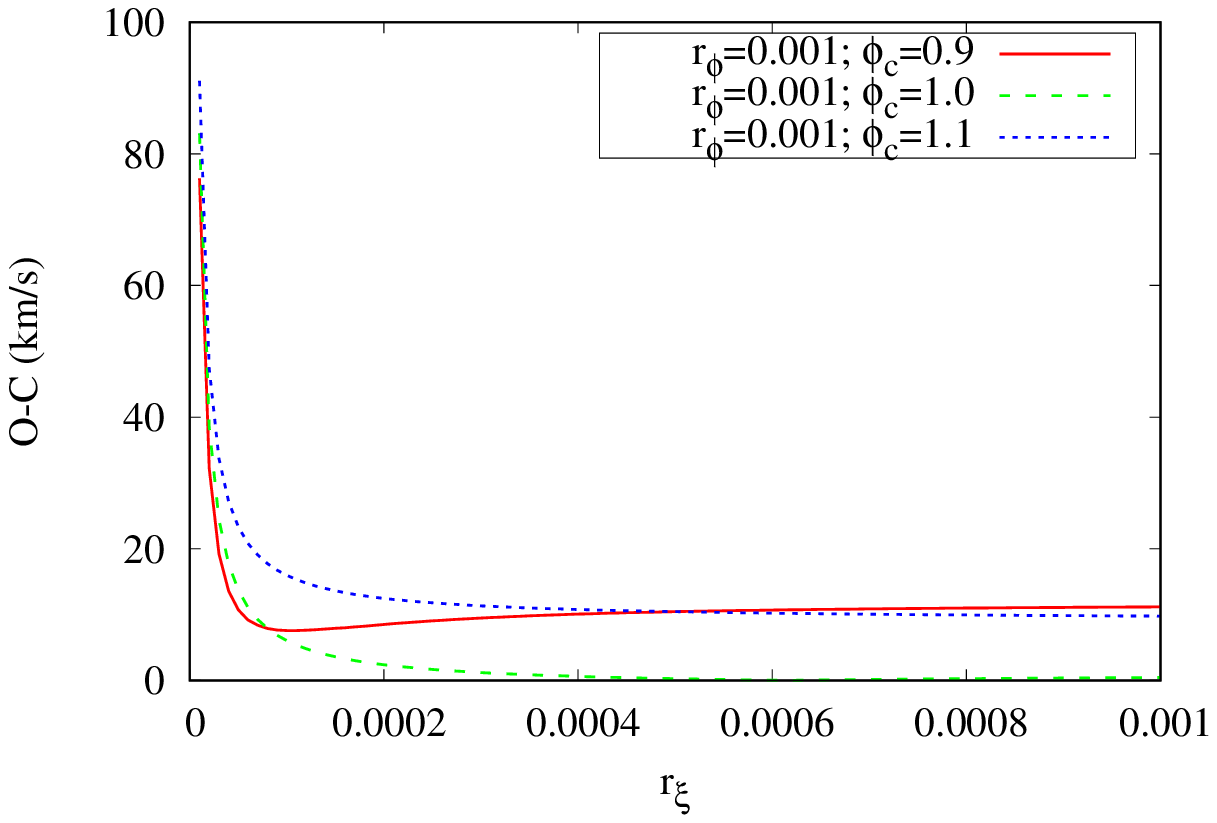}
\includegraphics[width=0.45\textwidth]{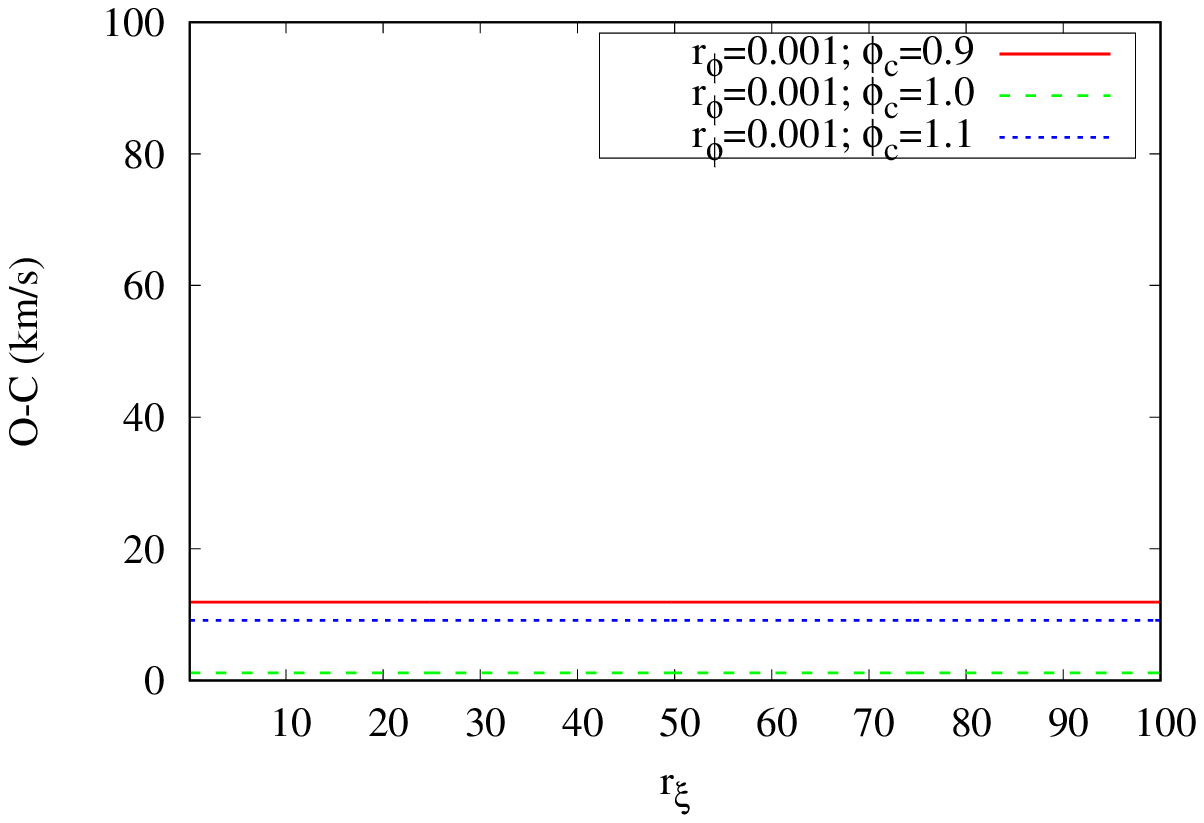}
\caption{The difference between observed and calculated values for circular velocity (O-C) (km/s) of elliptical galaxies: upper left: case of parameter $\phi_c$ ($r_{\xi}$: 0.0001 and 0.001 kpc; $r_{\phi}$: 0.0001 and 0.001 kpc), upper right: case of parameter $r_{\phi}$ ($r_{\xi}$: 0.001 kpc; $\phi_c$: 0.9, 1.0 and 1.1); down left: case of parameter $r_{\xi}$ ($r_{\phi}$: 0.0001 kpc; $\phi_c$: 0.9, 1.0 and 1.1), down right: the same like down left, but for higher values of parameter $r_{\xi}$.}
\label{fig06}
\end{figure*}

\begin{figure*}[ht!]
	\centering
	\includegraphics[width=0.70\textwidth]{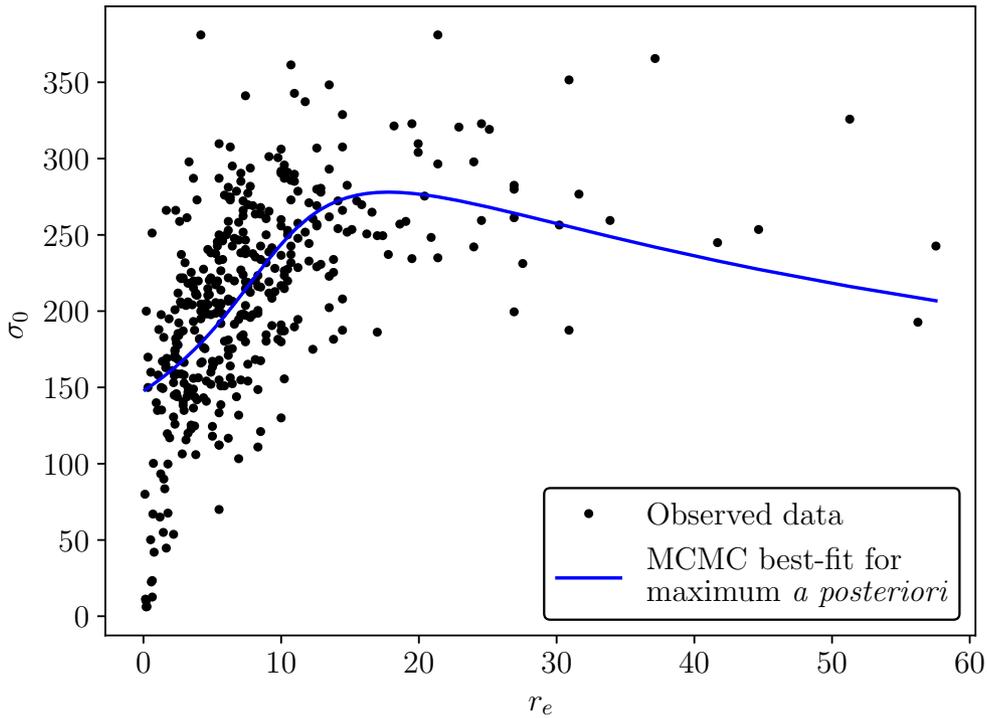}
	\caption{MCMC best-fit of $\sigma^{theor}(r_e)$ (blue solid line) to the observations (black circles). The best-fit was obtained for maximum \textit{a posteriori}, corresponding to the 50$^{th}$ percentile of the posterior probability distributions.}
	\label{fig07}
\end{figure*}

\begin{figure*}[ht!]
	\centering
	\includegraphics[width=0.95\textwidth]{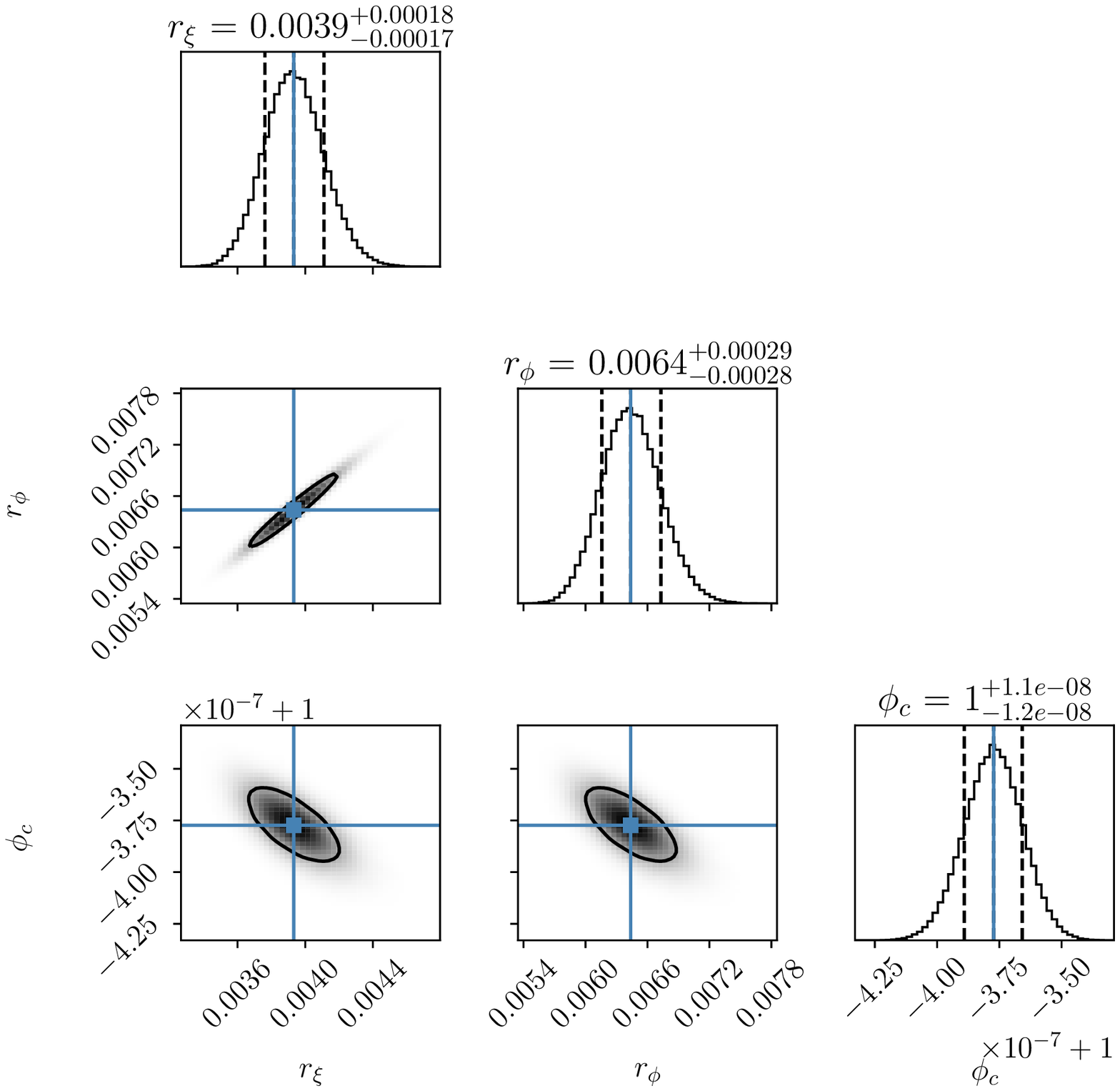}
	\caption{The posterior probability distributions of the parameters of Non-local gravity model. The contours reported 68\% confidence level for all studied parameters.}
	\label{fig08}
\end{figure*}

\section{Data}

In order to compare theoretical results with observations, we use data reported in Table I of Ref. \citet{burs97}. 
These data are the result of several observational campaigns
over the years \citet{burs97}. From Table 1 we used
effective radii, effective luminosities and characteristic velocities of galaxies, galaxy groups, galaxy clusters and globular clusters. 

For our investigation, we use columns (5), (6), (7) and (8) of Table 1 (see Appendixes A) and paper \citet{burs97}, as well as the notation for circular velocity $v_c$ from that table: for ellipticals, it is $v_c$ = $\sigma_0$. 

In order to better explain the data we are using, here we give Figures \ref{fig01} and \ref{fig02}. The total number of galaxies is 1150, while among them there are 400 elliptical galaxies. We show galaxy masses $M_e$ as a function of effective radius $r_e$, for elliptical and other galaxies (Fig. \ref{fig01}), and also galaxy luminosities $L_e$ as a function of $r_e$ (Fig. \ref{fig02}). From right panel in both figures we can see these properties of galaxies, for locations closer to the center i.e. for smaller values of $r_e$.

\section{Results and discussion}

We study velocity dispersion $\sigma^{theor}$ as a function of the effective radius $r_e$. Specifically, we constrain the parameters of Non-local Gravity using a sample of elliptical galaxies given in \citet{burs97}. Taking into account that the effective radii of some of the largest known elliptical galaxies, such as e.g. the supergiant IC 1101, are on the order of 50 kpc \citep[see e.g. Table 2 in][]{fish95}, we assumed the values of the characteristic radii of Non-local Gravity $r_\xi$ and $r_\phi$ ranging from very small ones up to the several hundred kpc (see \ref{fig03}--\ref{fig06}). In order to check how different kinematical properties influence the Non-local Gravity parameters, we investigate the gravitational parameters according to the following values: $r_{\xi}$: 0.001, 0.01, 0.10, 1.00 kpc and $r_{\phi}$: 0.001, 0.01 and 0.1 kpc. For the values of $\phi_c$, we considered values: 0.9, 1.0 and 1.1. One should have in mind that, in the observed sample, $\sigma_0$ is equal to the Newtonian circular velocity. The method that we are using is described in detail in references \citet{bork16,bork18,capo20a,bork21} and references therein.

Fig. \ref{fig03} shows velocity dispersion $\sigma^{theor}$ as a function of the effective radius $r_e$ for elliptical galaxies, for four different values of the $r_{\xi}$: 0.001, 0.01, 0.10 and 1.00 kpc. Theoretical values of velocity dispersion $\sigma^{theor}$ are calculated for the three values of Non-local Gravity parameter $r_{\phi}$: 0.001, 0.01 and 0.1 kpc. We study the case when values of $\phi_c$ is 0.9. We can see that the agreement between theoretical values and astronomical data is not  good. We can notice that theoretical value for $\sigma^{theor}$ is lower than values of corresponding observed data. A little better agreement is achieved in case when $r_{\xi}$ takes values near 0.001 kpc and $r_{\phi}$ takes values near 0.01 kpc. We study also the cases when values of $\phi_c$ are 0.7 and 0.8 and agreement with observations is more poor compared to case 0.9. 

Fig. \ref{fig04} represents the same as in Fig. \ref{fig03} but for the value of $\phi_c$ equals 1.0. Agreement is very good for all studied values of $r_{\xi}$ (from 0.001 to 1.0 kpc) and $r_{\phi}$ (from 0.001 to 0.1 kpc). The first term in Eq.\eqref{equ12} is dominant and is equals to the Newtonian one. In this case, the second, and especially the third term, are very small. Also, our analysis has shown that for larger values of Non-local Gravity parameters $r_{\xi}$ $>>$ 1 kpc and $r_{\phi}$ $>>$ 1 kpc, good agreement is achieved when parameter $\phi_c$ is very close to 1 (see Fig. \ref{fig06}). This result is expected because in that case the Non-local Gravity potential is close to the Newtonian limit, see  Eq.\eqref{equ04}.

Fig. \ref{fig05} represents the same as in Fig. \ref{fig03} but for the value of $\phi_c$ equals 1.1. The agreement between theoretical values and astronomical data is poor. A little better agreement is achieved in case when $r_{\xi}$ takes values near 0.01 kpc and $r_{\phi}$ takes values near 0.001 kpc. We can notice that theoretical value for $\sigma^{theor}$ is higher than values of corresponding observed data. We study also the cases when values of $\phi_c$ are 1.2 and 1.3. The agreement between theoretical values and astronomical data becomes more poor when difference ($\phi_c - 1$) increases by magnitude. Our analysis shows (see Fig. \ref{fig06}), that differences between observed and calculated values of circular velocity (O-C) are very sensitive to value $\phi_c$. In order to achieve better agreement, value of $\phi_c$ should be very close to 1.

Fig. \ref{fig06} represents the comparison between the observed (O) and calculated (C) velocities of elliptical galaxies (O-C) (km/s) in case of some specific values of Non-local Gravity parameters $\phi_c$, $r_{\xi}$ and $r_{\phi}$. By inspection of Fig. \ref{fig06} we can see that observed and calculated value for circular velocity is very sensitive on values od parameter $\phi_c$ and better agreement between observed and calculated velocities is obtain when $\phi_c$ is approach to 1, i.e. value (O-C) tends to be very small by magnitude. In case of parameters $r_{\xi} < $ 0.0001 kpc and $r_{\phi} <$ 0.0002 kpc value (O-C) has strong increase. These will affect that observed and calculated value for circular velocity will be very different. That's why these region of parameters $r_{\xi}$ and $r_{\phi}$ should be excluded. For values of parameters $r_{\xi}$ $\gtrsim$ 1 kpc and $r_{\phi}$ $\gtrsim$ 1 kpc, difference (O-C) has almost constant value. It means that varying of $r_{\xi}$ and $r_{\phi}$ will not influence $\sigma^{theor}$, and if $\phi_c$ is very close to 1, agreement between observed and calculated velocities will be satisfactory (in that case the Non-local Gravity potential given by Eq.\eqref{equ04} is very close to the Newtonian limit).

After inspection of Figs. \ref{fig03}--\ref{fig06}, we can conclude that the best agreement between theoretical results and astronomical observations is for values of $\phi_c$ very close to 1 and when both $r_{\xi}$ and $r_{\phi}$ take values between 0.001 and 0.01 kpc. Also, for larger values of Non-local Gravity parameters $r_{\xi} \wedge r_{\phi}$ $>>$ 1 kpc, satisfactory agreement is achieved only when parameter $\phi_c$ is very close to 1 (i.e. to the Newtonian limit).

\subsection{Uncertainty analyses of the gravity parameters using Markov chain Monte Carlo analysis}

Probabilistic data analysis has great impact on scientific research in the past decade. Its procedures involve computing and using the posterior probability density function for the parameters of the model or the likelihood function. Markov chain Monte Carlo (MCMC) methods provide sampling approximations to the posterior probability density function efficiently even in parameter spaces with large numbers of dimensions. For example, many problems in cosmology and astrophysics have benefited from MCMC because there are many free parameters, and the observations are usually low in signal-to-noise ratio \citep{fore13}. 
	
In this subsection we estimated 68\% confidence region for the Non-local Gravity parameters using MCMC \citep[for more details see][and references therein]{fore13,audr13,shar17,hogg18}. For that purpose, we used an MIT licensed pure-Python implementation of Goodman \& Weare's Affine Invariant Markov chain Monte Carlo Ensemble sampler (\url{https://emcee.readthedocs.io/en/stable/}). The explanation of the \texttt{emcee} algorithm and its implementation in detail are given in the following paper: \citet{fore13}.

As a first step, we obtained the maximum likelihood values of Non-local Gravity parameters using the \texttt{optimize.minimize} module from SciPy for maximization of their likelihood function (or more precisely, for minimization of its negative logarithm), assuming Eq. (\ref{equ14}) as our model for $\sigma^{theor}$. After that we used these maximum likelihood values of the parameters as a starting point for our MCMC simulations, which we performed in order to estimate the posterior probability distributions for the gravity parameters. These simulations were carried out using $0 < r_\xi \wedge r_\phi < 100$ kpc and $0.9 < \phi_c < 1.1$ as our priors.

Fig. \ref{fig07} represents comparison between the MCMC best-fit of $\sigma^{theor}(r_e)$ (blue solid line) with observations (black circles). This MCMC best-fit is obtained for maximum \textit{a posteriori} \citep[see][for more details]{hogg10} which corresponds to the 50$^{th}$ percentile of the posterior probability distributions. Errors for data $\sigma_0$ were not given in Table I of Ref. \citet{burs97}. We assumed them to be equal and fixed to $\sigma_{0,err}$ = 25 km/s, obtained by roughly averaging the uncertainties presented in the following observational studies: \citet{nell95,fabr13,saul13}. 

Fig. \ref{fig08} represents the obtained posterior probability distributions of the parameters of Non-local Gravity model ($r_{\xi}$, $r_{\phi}$, $\phi_c$), where the contours represent their 68\% confidence levels. The best-fit values of Non-local Gravity parameters and their uncertainties, obtained from the 16$^{th}$, 50$^{th}$ and 84$^{th}$ percentiles of the posterior probability distributions, are: $r_{\xi} =0.0039^{+0.00018}_{-0.00017}$ kpc, $r_{\phi} = 0.0064^{+0.00029}_{-0.00028}$ kpc and $\phi_c = 1^{+1.1e-08}_{-1.2e-08}$.

We can conclude that results obtained by MCMC simulation (see Figs. \ref{fig07} and \ref{fig08}) are consistent with our previous analyses using Figs. \ref{fig03}--\ref{fig06}, but additionally give us the best-fit values of parameters of Non-local Gravity model.

\section{Conclusions}

Non-local Gravity is a motivated approach at IR scales like cosmology. It is capable of giving a good explanation of late-time cosmological dynamics without the need of exotic forms of matter-energy. However, a theory of gravity should be valid at any scale. Here, we considered Non-local Gravity at galactic scales. Specifically, we investigated the velocity distribution of elliptical galaxies in the framework of a Non-local Gravity model without using dark matter hypothesis. We constrain its parameters using a sample of elliptical galaxies given in \citet{burs97}.

We can conclude that the best agreement between theoretical results and astronomical observations is achieved when both $r_{\xi}$ and $r_{\phi}$ are between 0.001 and 0.01 kpc. For large values of $r_{\xi}$ $>>$ 1 kpc and $r_{\phi}$ $>>$ 1 kpc good agreement is achieved only when parameter $\phi_c$ is very close to 1, because in that case the Non-local Gravity potential is very close to its Newtonian limit. Using MCMC model we obtained the following values and uncertainties of Non-local Gravity parameters: ($r_{\xi} =0.0039^{+0.00018}_{-0.00017}$ kpc, $r_{\phi} = 0.0064^{+0.00029}_{-0.00028}$ kpc and $\phi_c = 1^{+1.1e-08}_{-1.2e-08}$).

As a final remark, it is worth noticing that non-local contributions in the Einstein-Hilbert action can both solve shortcoming related to quantum effects in curved spacetimes, at fundamental UV scales, and can also result as characteristic lengths at IR regime. Such lengths have the potentiality to naturally address galactic dynamics, large scale structure, as well as accelerated cosmic expansion. In this perspective, their investigation could be extremely advantageous from an observational point of view.

\section*{Acknowledgments}
This work is supported by Ministry of Education, Science and Technological Development of the Republic of Serbia. P.J. wishes to acknowledge the support by this Ministry through the Project contract No. 451-03-68/2022-14/200002. SC acknowledges the support of {\it Istituto Nazionale di Fisica Nucleare} (INFN), {\it iniziative specifiche} MOONLIGHT2 and QGSKY.

\section*{Appendix A: Table 1 from \citet{burs97}}
Table 1 in \citet{burs97} lists the data (the first page only is printed for the journal). The table contains effective radii, effective luminosities and characteristic velocities which have been derived for gravitationally-bound stellar systems, including galaxies, galaxy groups, galaxy clusters and globular clusters. Also, related data, including stellar population information, are included, as summarized in the table explanation \citet{burs97}.

The full table is electronically available in a convenient ASCII format given in 'metaplanetab1' among 'source' files of its arXiv version: \url{https://arxiv.org/e-print/astro-ph/9707037}. The data are organized in 19 columns there, and for our investigation the most important were the following ones:
\begin{itemize}[nosep,label=-]
	\item Column 1 gives the name for the object. 
	\item Column 2 gives an internal identifying number. 
	\item Column 3 gives a numerical code for the type of object.
	\item Column 4 gives the distance in units of Mpc. 
	\item Column 5 gives the logarithm of the original characteristic internal velocity.
	\item Column 6 gives either the observed value of $\log r_c$ or the transformed value $\log \sigma_0$. 
	\item Column 7 gives the values of $\log r_e$ in kpc.
	\item Column 8 gives the values of $\log I_e$ in $L_0$ pc$^{-2}$.
	\item Columns 15 and 16 give effective mass $\log M_e$ and effective B-band luminosity $\log L_e$, in solar units.
\end{itemize}

%% Bibliography
\bibliographystyle{model5-names}
\biboptions{authoryear}
\bibliography{literatura}

\end{document}